# Binary Multifunctional Ultrabroadband Self-Powered g-C$_3$N$_4$/Si Heterojunction High-Speed Photodetector

*Nisha Prakash, Gaurav Kumar, Manjri Singh, Arun Barvat, Prabir Pal,[*] Surinder P. Singh, H. K. Singh, Suraj P. Khanna[*]*

Nisha Prakash, Gaurav Kumar, Manjri Singh, Arun Barvat, Dr. Prabir Pal, Dr. Surinder P. Singh, Dr. Suraj P. Khanna
CSIR-National Physical Laboratory, Dr. K. S. Krishnan Road, New Delhi 110012, India
E-mail: palp@nplindia.org
E-mail: khanna@nplindia.org

Nisha Prakash, Gaurav Kumar, Manjri Singh, Arun Barvat, Dr. Prabir Pal, Dr. H. K. Singh
Academy of Scientific and Innovative Research, CSIR-National Physical Laboratory (Campus), Dr. K. S. Krishnan Road, New Delhi 110012, India

Keywords: binary, g-C$_3$N$_4$, hybrid, energy harvesting, photosensitivity

Abstract: Compact optical detectors with fast binary photoswitching over a broad range of wavelength are essential as an interconnect for any light-based parallel, real-time computing. Despite of the tremendous technological advancements yet there is no such single device available that meets the specifications. Here we report a multifunctional self-powered high-speed ultrabroadband (250-1650 nm) photodetector based on g-C$_3$N$_4$/Si hybrid 2D/3D structure. The device shows a novel binary photoswitching (change in current from positive to negative) in response to OFF/ON light illumination at small forward bias ($\leq 0.1$ V) covering 250-1350 nm. At zero bias, the device displays an extremely high ON/OFF ratio of $\sim 1.2 \times 10^5$ under 680 nm (49 µWcm$^{-2}$) illumination. The device also shows an ultrasensitive behaviour over the entire operating range at low light illuminations, with highest responsivity (1.2 AW$^{-1}$), detectivity (2.8 $\times$ 10$^{14}$ Jones) and external quantum efficiency (213%) at 680 nm. The response and recovery speeds are typically 0.23 and 0.60 ms, respectively, under 288 Hz light switching frequency. Dramatically improved performance of our device is attributed to the heterojunctions formed by the ultrathin g-C$_3$N$_4$ nanosheets embedded in the Si surface.

Polymeric graphitic carbon-nitride (g-$C_3N_4$) is an emerging two-dimensional (2D) layered material which is structurally congruent to graphene. It primarily constitutes of π-conjugated graphitic planes formed via $sp^2$ hybridized carbon (C) and nitrogen (N) atoms in an alternating fashion in the basic graphene lattice with strong covalent in-plane bonds while the interlayers are held together by weak van der Waals forces.[1] The major building block of g-$C_3N_4$ is termed as tri-s-triazine rings, which is $C_6N_7$ connected through planar tertiary N bridges. In contrast to graphene, g-$C_3N_4$ offers potential for semiconductor device applications as it has a suitable band gap of ~2.7 eV. Bulk g-$C_3N_4$ has been shown as a useful material for visible light absorption and solar cell applications.[2-4] Despite of the fascinating properties of bulk g-$C_3N_4$, the reduced dimensionality of g-$C_3N_4$ nanosheets have drawn increasing attention conceivably due to enhanced light absorption below 450 nm, large specific surface area, high electrical conductivity and efficient separation of electron-hole pairs. It is also established that an increase in the bandgap is observed in the 2D nanosheets compared with that of the bulk counterpart due to the quantum confinement effects.[5] However, self-doping in g-$C_3N_4$ nanosheets increases the electrical conductivity with extended light absorptions.[6] The ultrathin g-$C_3N_4$ nanosheets have successfully demonstrated their potential in photocatalysis,[7, 8] bio-imaging,[9] $CO_2$ conversion,[10] air purification[11] and novel photovoltaic applications.[12, 13] Further, Zhang *et al*. have reported an increased density of states in the conduction band as compared to its bulk counterpart through the first-principle calculations and thereby an improved optical and electronic response in g-$C_3N_4$ nanosheets.[9]

Interestingly, novel ultrathin organic-inorganic material integration has led to several hybrid nanostructures and nanocomposites for enhanced photodetection applications utilizing a range of different 2D and 3D materials.[14, 15] Specifically, g-$C_3N_4$ hybrid thin film photodetector consisting of inorganic $MoS_2$ and organic g-$C_3N_4$ nanosheets forming 2D/2D heterojunction hybrid device was demonstrated for ultraviolet and visible photodetection by Velusamy *et al*.[16] Similarly, Lai *et al.* have reported a g-$C_3N_4$/graphene based hybrid phototransistor that showed a high responsivity, but its sensitivity was limited to UV-A detection.[17] Lublow *et al*. reported two-step HF etching method firstly to modify the silicon surface, secondly to under etch the g-$C_3N_4$ deposited on the Si substrate leading to nanosheets or lamellae embedded into the Si substrate for photocatalytic applications. The fabricated hybrid structure displayed rectifying heterojunction behavior, enhancing the photogenerated electron conduction towards the g-$C_3N_4$/Si interface.[18]

Recently, a number of low powered hybrid photodetectors have been demonstrated displaying a remarkable feature where the photoswitching resembles ″0″ and ″1″ of a binary digital code making them suitable for weak signal detection and light-based computing.[19-21] However, the binary switching behavior of such devices reported till date is limited to UV and visible spectral range. In the present paper, we report the study on multifunctional hybrid photodetector based on ultrathin g-$C_3N_4$ nanosheets embedded into the Si surface. The device displayed an extended ultrabroadband photodetection in self-powered mode from 250 to 1650 nm. The device operates with high speed over the entire spectrum range covering ultraviolet-visible (UV-Vis) and near-infrared (NIR) range. In addition, the device demonstrates a binary response at small forward bias over a wide spectral range of 250 to 1350 nm, which makes it suitable for light-based computing. Such a wide range of binary switching behavior has not been reported. Our device exhibited an excellent photodetection performance in terms of ON/OFF ratio, spectral photosensitivity, detectivity, and response and recovery times. This remarkably low-cost and highly versatile ultrabroadband hybrid photodetector may provide a novel route for monolithic integration of 2D and 3D materials.

**Figure 1**a schematically illustrates the exfoliation of the bulk g-$C_3N_4$ powder into thin nanosheets upon ultrasonication in the ethanol solution for different exfoliation time periods of 2, 6 and 12 h.[22] The size distribution of as-exfoliated g-$C_3N_4$ nanosheets as a function of sonication time was investigated by atomic force microscopy (AFM) image (Figure 1b). It is evident that the bulk g-$C_3N_4$ has been successfully exfoliated into nanosheets. The estimated

lateral size and thickness of the exfoliated nanosheets are listed in the Figure. One can see that the nanosheets thickness and sizes gradually decreases as we increase the sonication time. It can be seen that the lateral size of the nanosheets ranges from 50 to 100 nm with an average thickness of the order of ~1-2 nm, indicating the gCN-12 is comprised of only about two to six C-N layers. The molecular structure was investigated by Fourier transform infra-red spectroscopy (FTIR) in the range of 500 to 4000 cm$^{-1}$ displayed in Figure 1c. The FTIR peaks observed from 800 to 880 cm$^{-1}$ indicate the characteristic breathing mode of s-triazine units, corresponding to the condensed C-N heterocycles. The peaks appearing between 1200-1700 cm$^{-1}$ are assigned to the stretching vibration modes of C-N heterocycles. The N-H group related features observed between 2900-3700 cm$^{-1}$ were found to increase with increasing edge boundaries in highly cleaved and exfoliated nanosheets.[8, 23] In addition to this, the peak observed at 1250 cm$^{-1}$ due to C–H appeared only after 8 h sonication (not shown) and further the intensity became pronounced with sonication and could be ascribed to breaking of sheets from bridging N atoms. The appearance of IR peak at 1640 cm$^{-1}$ due to C=C/C=N increases with sonication, suggesting the sheets are highly exfoliated. FTIR spectra indicate that the graphitic nature in the nanosheets is maintained even after 12 h sonication with enhanced specific surface area of gCN-12.

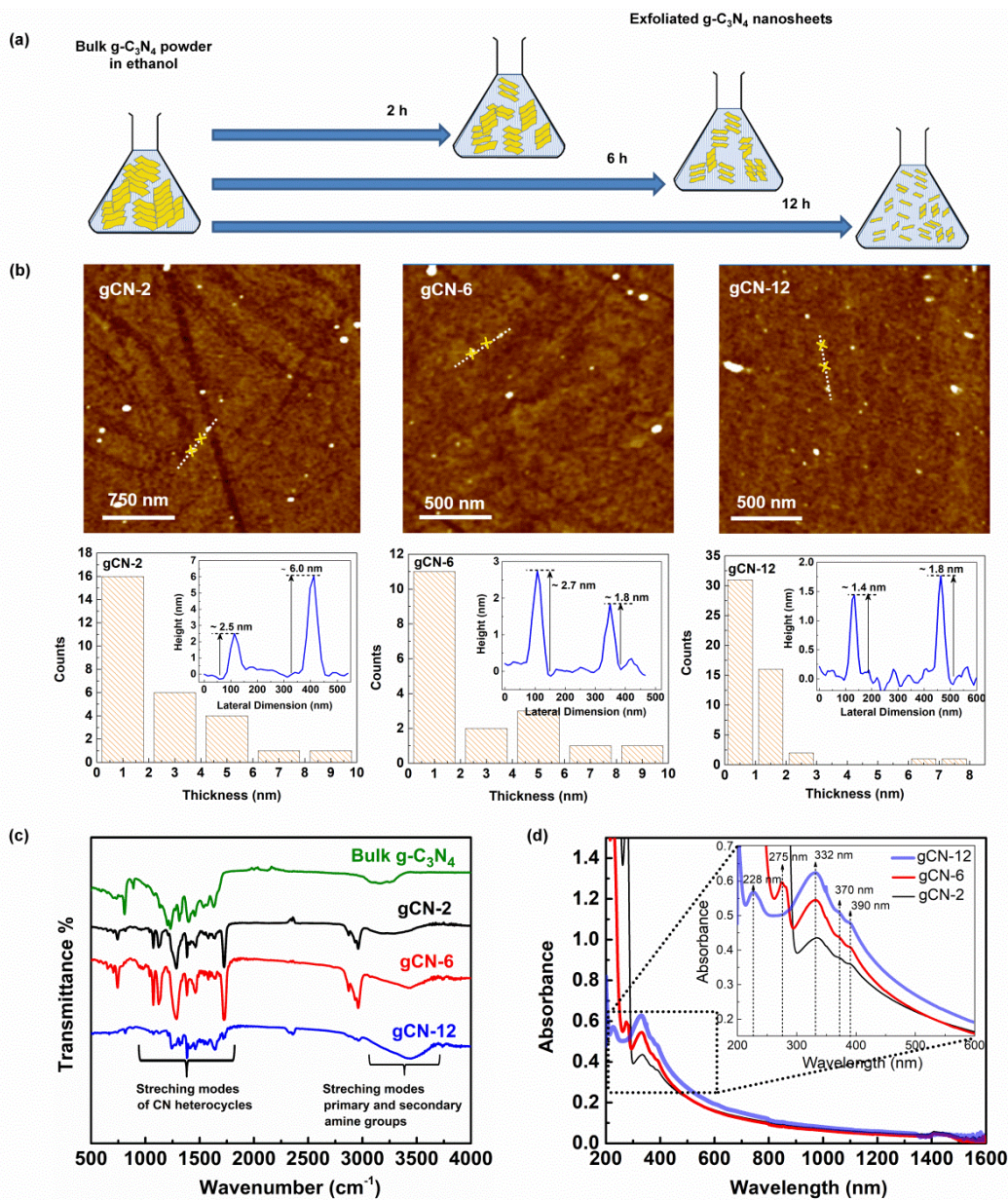

**Figure 1.** a) Schematic illustration of the mechanism of exfoliation upon sonication, where the nanosheets are exfoliated for three time durations, 2 h (gCN-2), 6 h (gCN-6), and 12 h (gCN-12). b) Typical AFM images of gCN-2, gCN-6 and gCN-12 nanosheets randomly dispersed on etched p-Si surface along with respective line profile of the sheets c) FTIR spectra of the bulk g-$C_3N_4$ powder and the exfoliated gCN-2, gCN-6 and gCN-12. d) Typical UV-Vis-NIR spectra of exfoliated colloidal solutions of bulk, gCN-2, gCN-6 and gCN-12.

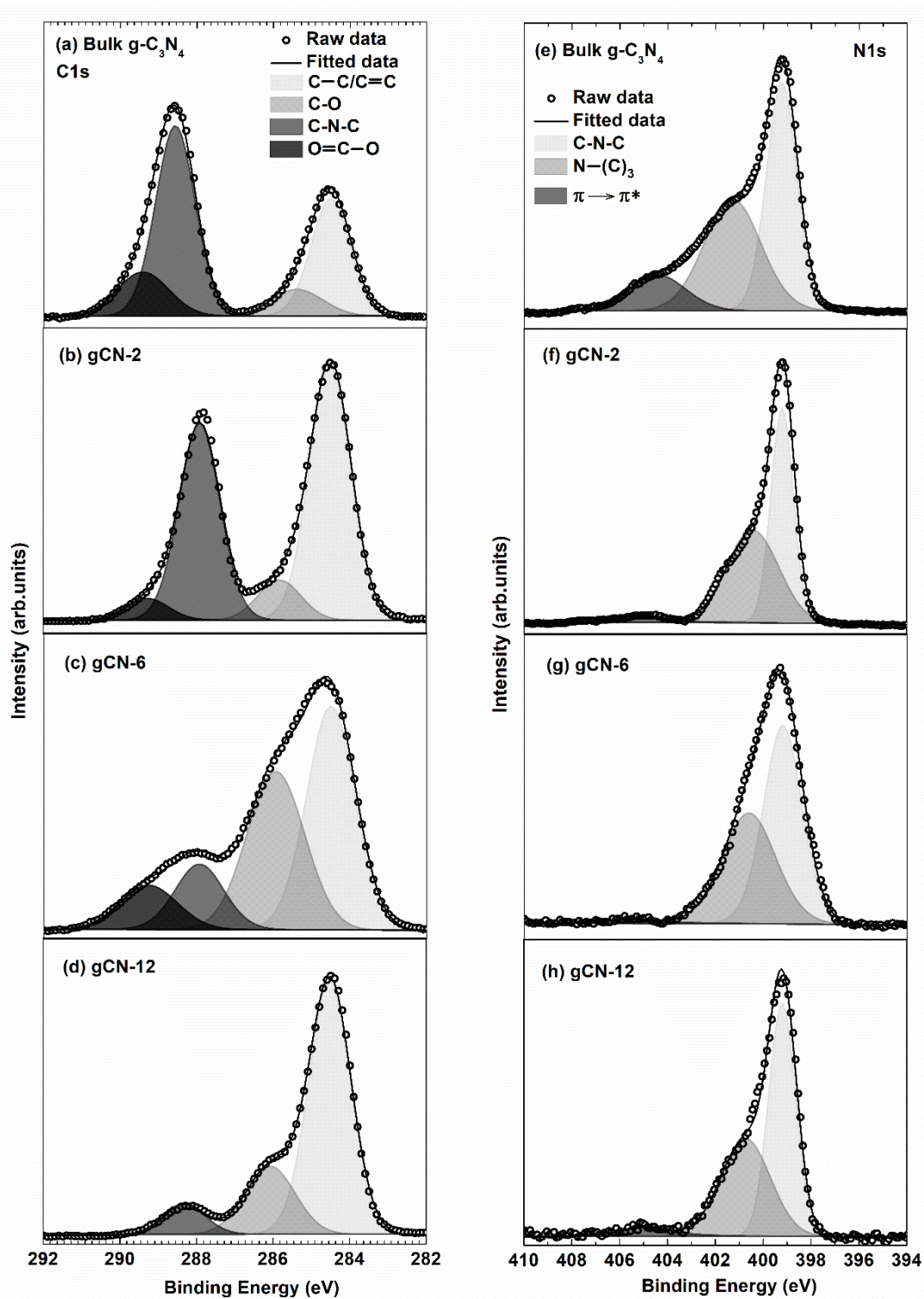

**Figure 2.** High-resolution XPS spectra corresponding to C 1s (left panels) and N 1s spectra (right panels) for as-prepared exfoliated nanosheets drop-casted on p-Si substrates and also bulk g-C$_3$N$_4$ powder. The spectra highlight the significant variation in carbon and nitrogen species present on the gCN-12 surfaces.

The optical absorption properties of the exfoliated g-C$_3$N$_4$ samples are shown in Figure 1d. The integrated area of the absorption band between 200 to 800 nm for gCN-2, gCN-6 and

gCN-12 samples were found to be 110, 130 and 160, respectively. This confirms a distinct increase in the surface area as we increase the exfoliation time. An absorption edge was observed near 450 nm, consistent with the band gap (~2.75 eV) reported earlier.[24] Although the g-$C_3N_4$ related transitions at ~332 nm along with two weak features at ~370 and ~390 nm remain unchanged for all three samples, the onset absorption peak is UV shifted at higher exfoliation times. The peaks at 332, 370 and 390 nm are assigned to the $\pi \rightarrow \pi^*$ transitions. For gCN-12, the onset of the absorption spectrum is shifted to the deep UV region with suppression in intensity compared with that of the other samples. It is known that in g-$C_3N_4$ ring, the lone-pair (LP) states are formed within the $\pi$ band for $sp^2$ N, while it forms within $\pi$ and $\pi^*$ bands for $sp^3$ N.[25] Owing to the presence of both $sp^2$ and $sp^3$ N in g-$C_3N_4$ ring, the LP $\rightarrow \pi^*$ transition originated from $sp^3$ N is expected in our case. The observed blue shift with suppression of the LP $\rightarrow \pi^*$ transition for gCN-12 sample is due to decrease in $sp^3$ N.[26] The NIR absorption observed in the nanosheets is attributed to the delocalization of $\pi$ electrons in the layered nanosheets enhancing the out-of-plane mobility.

The oxidation states of the chemical constituents of the exfoliated nanosheets were determined by X-ray photoelectron spectroscopy (XPS). The high-resolution XPS spectra of C 1s and N 1s are shown in **Figure 2**. The bulk g-$C_3N_4$ powder sample was analyzed first to provide a benchmark for nanosheets. The C 1s spectrum from powder g-$C_3N_4$ was fitted with four components at binding energies of 284.5, 285.4, 288.5 and 289.4 eV. The peak at 284.5 eV is assigned to C−C, C=C and/or adventitious carbon commonly observed on surfaces by XPS.[27] The peak located at 288.5 eV is due to C-N-C coordination.[6] The additional two weak peaks at 285.4 and 289.4 eV is assigned to C-O species and ester/carboxylic groups (O=C–O), respectively.[28] Similarly, the spectrum from N 1s from powder sample taken as a benchmark for exfoliated samples was deconvoluted by three components at binding energies of 399.2, 401.2 and 404.4 eV. The main peak at 399.2 eV is assigned to C-N-C coordination, while the peak at 401.2 eV is assigned to sp$^3$ N−(C)$_3$ of the tertiary N from the bridges and the centre of the tri-s-triazine units. The weak shoulder component at 404.4 eV is due to $\pi \rightarrow \pi^*$ excitations.[8] **Table S1** in supplementary information highlights the various component in C 1s and N 1s region for different samples. The intensity ratio of N−(C)$_3$ to C-N-C coordination were found to be 0.55 and 0.42 for bulk and gCN-12 samples, respectively, suggesting self-doping of tertiary N from the bridges with C.[6] This happens simultaneously with an enhancement of C−C, C=C and/or adventitious carbon component at the surface of the nanosheets,[9, 27] consistent with our UV-Vis-NIR absorption spectra. These could be ascribed to the high electrical conductivity of the ultrathin nanosheets leading to improved optoelectronic properties.[6] We have studied many such samples and all of them have shown similar enhancement in electronic and opto-electronic behaviour. Henceforth the electrically active gCN-12 has been used for fabricating the heterojunction photodetector.

**Figure 3**a shows the schematic of the device. The room temperature current-voltage (I-V) characteristics recorded under dark and light illuminations from 350 to 1150 nm plotted in Figure 3b depicts typical higher photocurrent response in the reverse-bias mode. The observed asymmetric I-V characteristics under dark conditions as shown in the inset of Figure 3b demonstrate a typical rectifying behavior confirming the formation of an n-p junction between g-C$_3$N$_4$ and p-Si with an extremely low dark current of ~23 pA. Figure 3c shows I-V characteristics under dark and light illuminations beyond 1200 nm in semilog scale, depicting atypical behavior where a higher photo current response is observed in the forward-bias mode rather than the reverse-bias. The photoresponse observed below bandgap illuminations could be ascribed to the two-photon absorption causing it to absorb NIR photons by the occurrence of metastable states in the midgap region.[29, 30] Figure 3d demonstrates the variation of open-circuit voltages (V$_{OC}$) as a function of different light illuminations, wherein the highest V$_{OC}$ of 0.21 V was observed at 680 nm with an incident

power density of 49 µWcm$^{-2}$. A measurable photovoltaic effect is also observed for the illuminations from 1200 to 1650 nm displayed in the inset of Figure 3d. As per our knowledge, the observed $V_{OC}$ values over a wide spectral range, from UV to NIR, are among the best reported for silicon-based self-powered hybrid photodetectors.[31-33]

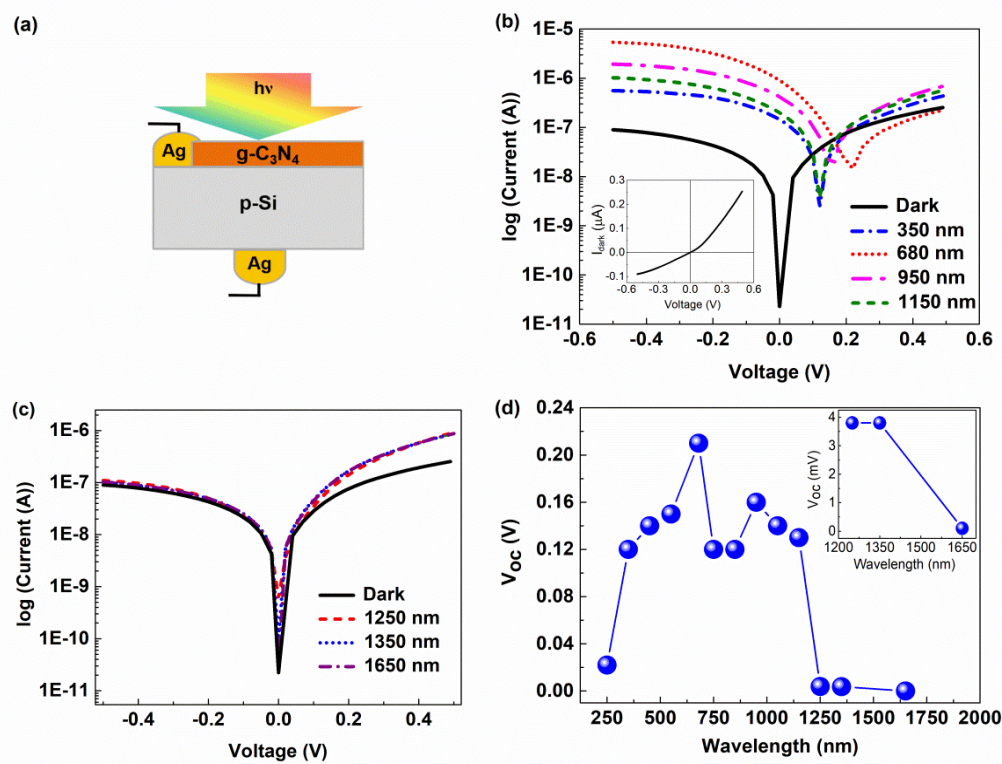

**Figure 3.** (a) Schematic diagram of the g-C$_3$N$_4$/Si hybrid photodetector. (b) The current-voltage (I-V) characteristics of the device under dark and illumination at wavelengths of 350, 680, 950 and 1150 nm in semilog scale, showing a typical photovoltaic behavior. The inset shows the dark I-V in linear scale. (c) I-V at dark and wavelengths of 1250, 1350 and 1650 nm in semilog scale. (d) Open-circuit voltage ($V_{OC}$) of the photodetector as a function of the wavelength. The inset highlights a measurable $V_{OC}$ beyond 1200 nm.

The photosensitivity, defined as $I_{ph}/I_{dark}$, where $I_{ph} = I_{light} - I_{dark}$, was measured at zero bias under 250 to 1650 nm wavelength range as shown in **Figure 4**a. The device shows very high photosensitivity (>10$^4$), in the entire wavelength length range from 350 to 1150 nm. The device exhibits highest photosensitivity (1.2 × 10$^5$) under 680 nm (49 µWcm$^{-2}$) wavelength illumination at zero bias. This is among the best values reported so far for heterojunction hybrid photodetectors.[21, 34] Further, the device also demonstrates good photosensitivity till near-infrared region. This confirms that our device can be projected as an ultrabroadband photodetector which can operate from 250 to 1650 nm under self-powered mode. In comparison to the commercial Si-DH photodetector, our device displayed an over all enhancement of photosensitivity ~550% in the spectral range of 250-1100 nm as shown in Supplementary **Figure S5**. Under bias, the observed photosensitivity is shown in Figure 4b. We must emphasize here that in the present case the bare p-Si, with similar silver contacts as the original device, did not show any photosensitivity under photovoltaic or photoconductive mode (shown in Supplementary **Figure S6**). The ON-OFF switching of the device at zero bias is shown in Figure 4c and 4d displaying stable and repetitive cycles illuminated over a wide spectral range, from UV to NIR. The detailed I-V characteristic curves along with transient photoresponse of the device under dark and light conditions from 350 to 1150 nm are shown in Supplementary **Figure S7**.

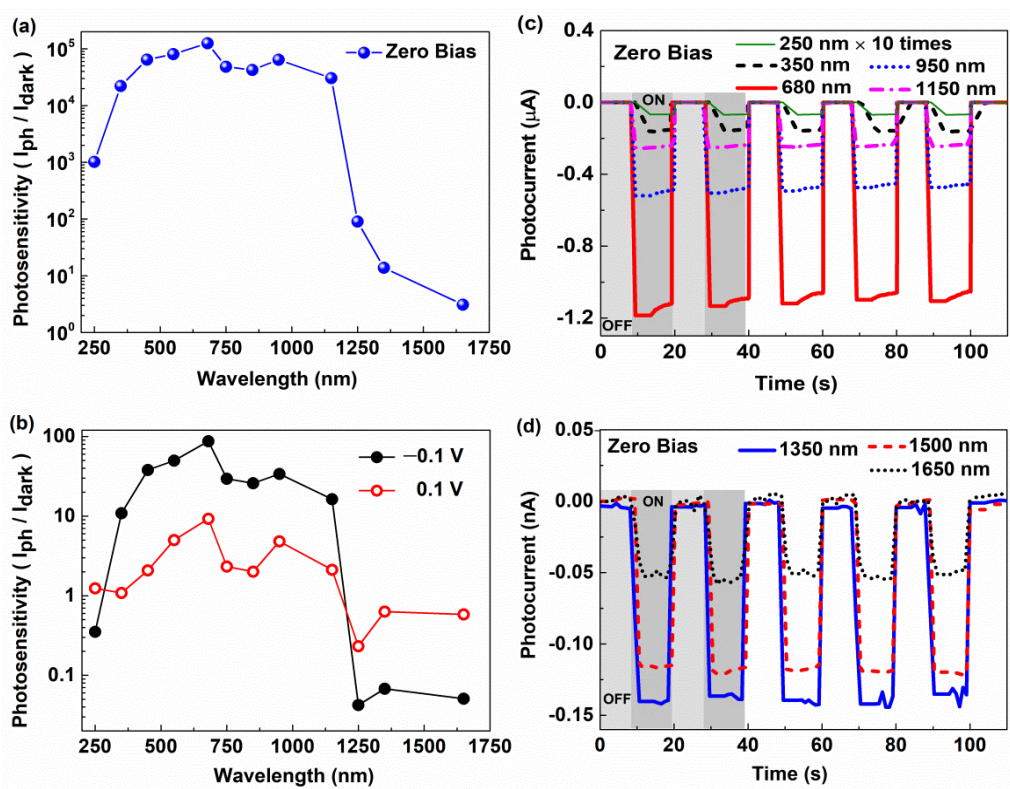

**Figure 4.** Photosensitivity measured under (a) zero bias, (b) small forward and reverse bias with light illuminations exhibiting ultrabroadband response from 250-1650 nm. Reproducible ON/OFF switching behavior of the device at zero bias (c) under light illuminations of 250, 350, 680, 950 and 1150 nm, (d) under illuminations of 1350, 1500 and 1650 nm.

The optical binary response is a key parameter of a photodetector required to realize the light-based computing. **Figure 5**a-g shows the binary response under small forward bias voltages over a wide spectral range. The current was observed to switch between positive (logic1) to negative (logic0) with light switching OFF to ON at small bias voltages between 250 to 1350 nm. In the forward bias condition, this sudden change in the polarity of current with light OFF to ON is observed when the number of photogenerated carriers exceeds the number of carriers injected by the external bias and thus reversing the net flow of current under the influence of built-in field. As soon as external forward bias exceeds the built-in voltage, the current flows in the direction of applied bias. This successfully demonstrates the ability of our device to be used as an optical binary switch operating for ultrabroadband (250-1350 nm) photodetection applications operating at room temperature, far exceeding the merits of high photosensitivities observed at zero bias.

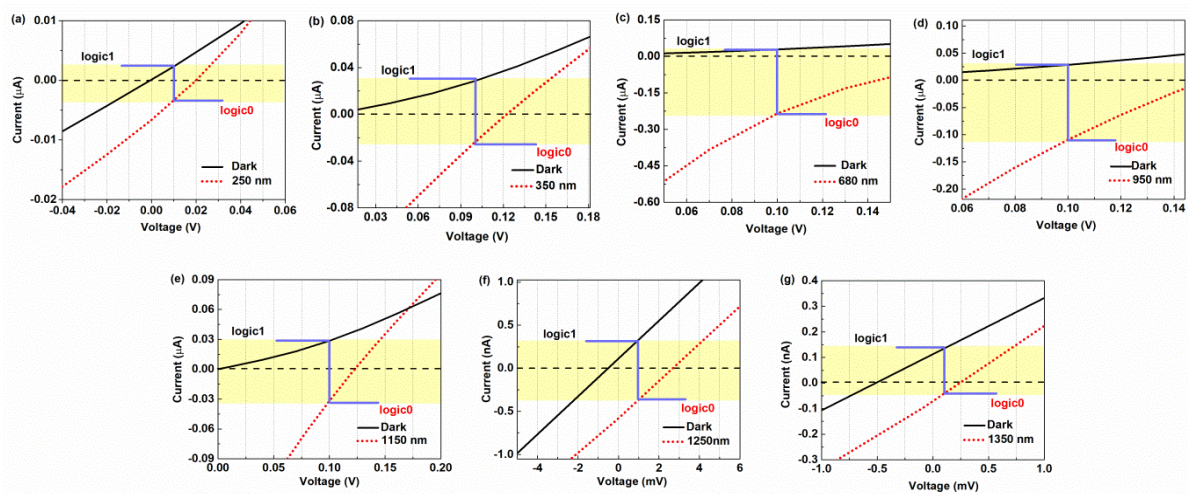

**Figure 5.** a-g) Shows the narrow range I-V curve in the dark (solid black) and upon various light illuminations (red dotted). The solid blue line is used to illustrate the binary behavior where current polarity changes from positive to negative in response to light OFF to ON, respectively, under fixed bias. The positive and negative current polarity represents high (logic1) and low (logic0) states of the device, respectively.

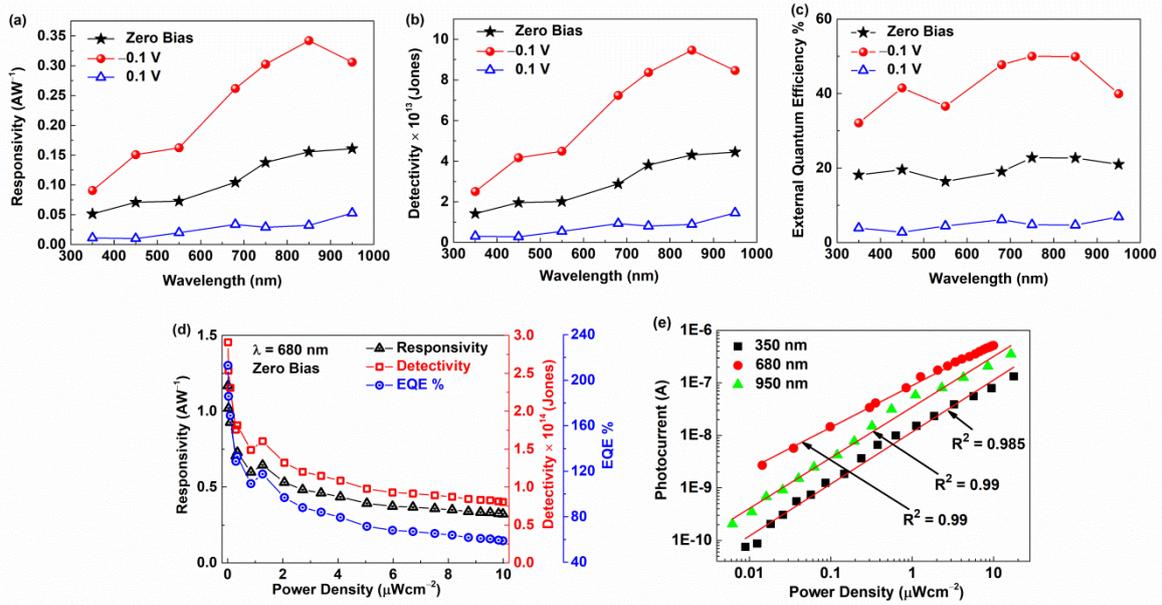

**Figure 6.** The response of the device as a function of wavelength, a) responsivity, b) detectivity and c) EQE. (d) Responsivity, detectivity and EQE of the device as a function of illumination intensity at wavelength 680 nm showing ultrasensitive behavior. (e) Photocurrent as a function of incident low light intensity measured with different wavelengths.

Furthermore, the performance of photodetectors is evaluated in terms of several figure of merit parameters, such as responsivity ($R_\lambda$), detectivity ($D^*$), external quantum efficiency (EQE (%)), response and recovery speed (for detailed formulas and calculation refer to Supplementary Information). **Figure 6**a-c show the responsivity, detectivity and EQE curves of the device as a function of wavelength. Under 850 nm light illumination with −0.1 V bias at a fixed intensity of 11 μWcm$^{-2}$, the calculated values of $R_\lambda$ = 0.34 AW$^{-1}$, $D^*$ = 9.46 × 10$^{13}$ cm-Hz$^{1/2}$W$^{-1}$ (Jones) and EQE = 50%, respectively. However, under zero bias the device displayed values of $R_\lambda$ (0.16 AW$^{-1}$), $D^*$ (4.3 × 10$^{13}$ Jones) and EQE (23%). A typical ultrasensitive behaviour under zero bias (680 nm) was observed at light intensity below 2 μWcm$^{-2}$ as shown in Figure 6d with $R_\lambda$ = 1.2 AW$^{-1}$, $D^*$ = 2.9 × 10$^{14}$ Jones and EQE = 213%. The details of the ultrasensitive behaviour of different wavelengths are provided in Supplementary Information as shown in **Figure S8**.

To investigate the recombination kinetics of charge carriers, photoresponse measurements were undertaken as a function of incident low light intensities. Figure 6e shows the photoresponse of the device under various incident power densities ranging from 0.01 µWcm$^{-2}$ to 10 µWcm$^{-2}$ operating at wavelengths 350, 680 and 950 nm. It was observed that the photocurrent shows a near to linear dependence with the incident light intensities useful for practical applications.

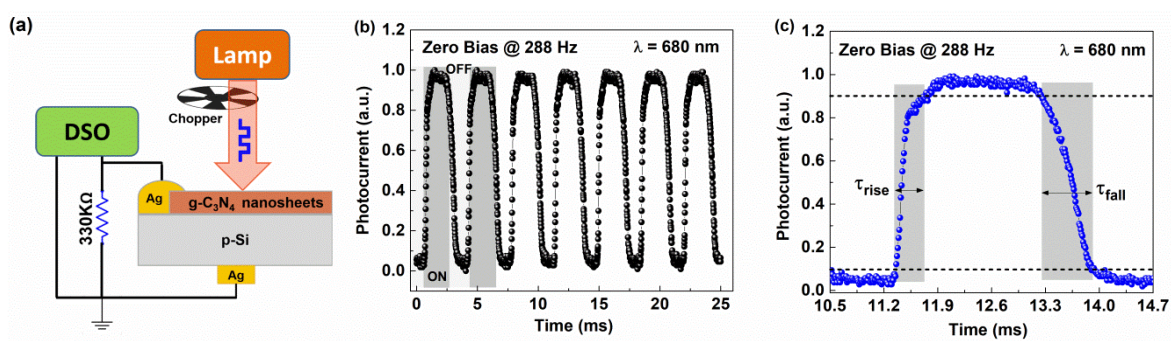

**Figure 7.** a) Shows the schematic diagram of our experimental setup to measure the speed of the photodetector. b) Normalized photocurrent response of the photodetector under a 288 Hz light switching frequency and c) a single normalized cycle displaying the response and recovery time.

**Table 1.** Shows the rise and fall time corresponding to 350, 680, 950 and 1150 nm.

| Wavelength [nm] | Rise time, $\tau_{rise}$ [ms] | Fall time, $\tau_{fall}$ [ms] |
|---|---|---|
| 350 | 0.40 | 0.62 |
| 680 | 0.23 | 0.60 |
| 950 | 0.13 | 0.43 |
| 1150 | 0.20 | 0.56 |

**Figure 7**a gives the schematic of the experimental setup used to measure the speed of the device. Figure 7b shows the transient photocurrent response of the device taken at zero bias under pulsed incident light of 680 nm chopped at 288 Hz. The response and recovery speed of a photodetector are generally assessed in terms of rise time ($\tau_{rise}$) and fall time ($\tau_{fall}$) as shown in Figure 7c. The rise time is defined as the time required for photocurrent to increase from 10% to 90% of maximum photocurrent and fall time is defined analogously. Further **Table 1**

gives the $\tau_{rise}$ and $\tau_{fall}$ measured at zero external bias under wavelengths from 350 to 1150 nm. These values are nearly 100 times faster than that of the recently reported hybrid photodetectors based on g-$C_3N_4$.[16, 17] The performance of our photodetector is compared with the best-reported values for other hybrid photodetectors and are summarized in **Table 2**. Furthermore, typical spectral response of commercial photodetectors are summarized in **Table 3**. In terms of device trade-off between spectral range, open-circuit voltage, speed, responsivity and detectivity, our hybrid photodetector demonstrates the best performance where a binary response is established. The utilization of 2D nanosheets integrated with p-Si in the hybrid device architecture enables the ultrabroadband photodetection below the bandgap of both the active materials.

**Table 2.** Comparison of the characteristic parameters for our hybrid photodetector and other reported hybrid photodetectors based on g-$C_3N_4$.

| Active material | Operating range [nm] | $I_{ON}/I_{OFF}$ ratio | Bias [V] | $R_\lambda$ [a)] [AW$^{-1}$] | $D^*$ [b)] [Jones] | $\tau_r / \tau_f$ [c)] [ms] | $I_{Dark}$ [d)] [pA] | REF |
|---|---|---|---|---|---|---|---|---|
| g-$C_3N_4$/p-Si | 350 (UV)<br>680 (Vis)<br>950 (NIR) | ~2.2 × 10$^4$<br>~1.2 × 10$^5$<br>~6.4 × 10$^4$ | 0 | ~0.05<br>~0.11<br>~0.16 | ~1.4 × 10$^{13}$<br>~2.9 × 10$^{13}$<br>~4.5 × 10$^{13}$ | ~0.40/0.62<br>~0.23/0.60<br>~0.13/0.43 | ~23 | This work |
| MoS$_2$/g-$C_3N_4$ | 365 (UV)<br>532 (Vis) | ~1 × 10$^4$<br>~4 × 10$^3$ | −9 | 4<br>0.7 | 8 × 10$^{10}$<br>4 × 10$^{11}$ | 50/80<br>60/95 | ~1 × 10$^3$ | Velusamy et al.[16] |
| g-$C_3N_4$ nanosheets/graphene | 370 (UV) | - | $V_G$ = 8<br>$V_{DS}$ = 0.5 | 4×10$^3$ | - | 15/121 × 10$^3$ | - | Lai et al.[17] |

a) Responsivity; b) Detectivity; c) Rise time / Fall time; d) Dark current

**Table 3.** Typical spectral response of commercial photodetectors.[35]

| Material | Wavelengths [nm] |
|---|---|
| Silicon | 400-1100 |
| Germanium | 600-1600 |
| GaAs | 400-900 |
| InGaAs | 900-1700 |
| InGaAsP | 800-1600 |

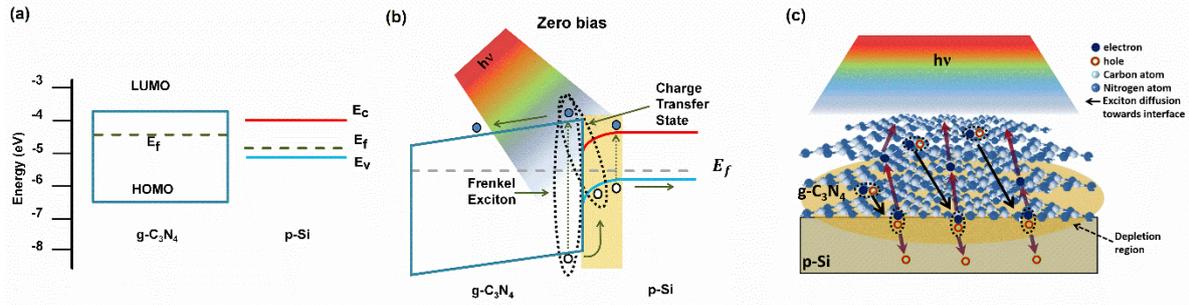

**Figure 8.** a) Equilibrium energy levels of g-$C_3N_4$ and p-Si, b) band diagram of g-$C_3N_4$ and p-Si interface with the movement of charge carriers under the influence of built-in field formed at the interface due to the difference in the work functions of g-$C_3N_4$ and p-Si. (c) Cross-section schematic showing diffusion of excitons towards the interface followed by dissociation and transport of charge carriers towards respective electrodes.

To gain more insight into the self-powered charge conduction mechanism, the energy level band diagram of the g-$C_3N_4$/p-Si hetero-interface can be drawn as shown in **Figure 8**. The equilibrium energy levels of g-$C_3N_4$ and p-Si are shown in Figure 8a. Here, the highest occupied molecular orbital (HOMO) and lowest unoccupied molecular orbital (LUMO) edges[36] of g-$C_3N_4$ are at −6.5 eV and −3.8 eV respectively, while the conduction band (CB) and valence band (VB) of p-Si are at −5.17 eV and −4.05 eV, respectively, with reference to vacuum level. Figure 8b shows the band alignment and the corresponding depletion region formation at the organic-inorganic hetero-interface of g-$C_3N_4$ and p-Si at zero bias. On illumination, excitons are generated in the depletion region on both sides of the interface as a result of photon absorption. The polymeric nature of g-$C_3N_4$ makes Frenkel type excitons[37] a possible candidate for photoexcited species in g-$C_3N_4$.[38] Frenkel type excitons are tightly bound electron-hole pairs with typical binding energies of 200 to 500 meV, much greater than the thermal energy at room temperature (25 meV) and thus are difficult to dissociate efficiently. In contrast to polymeric materials, weakly bounded Wannier type excitons are generated upon photo-absorption in highly crystalline inorganic semiconductors (p-Si) with binding energies of ~ 10 meV.[39] Here the HF-treated, passivated p-Si surface further helps to minimize the recombination centers by reducing the surface recombination velocity.[40] The energy offset between the LUMO of g-$C_3N_4$ and CB of p-Si is 0.25 eV whereas the offset is 1.33 eV between HOMO and VB of g-$C_3N_4$ and p-Si respectively, which enables the charge separation and transfer at the interface. The excitons formed near the junction in the g-$C_3N_4$ nanosheets migrate towards the interface where they form a hybrid charge transfer state which further rapidly dissociates into free carriers efficiently via built-in field at the interface. The energy loss involved in the formation of hybrid charge transfer states helps to reduce the higher binding energy of the Frenkel excitons and increases the probability of efficient separation of electrons and holes rather than recombination. After dissociation, the carriers in the g-$C_3N_4$ nanosheets may migrate independently via either of the three different types of conduction channels namely; intrachain, intraplanar or interplanar with higher contribution from the latter.[41, 42] The spatially separated individual photogenerated carriers transport through g-$C_3N_4$ and p-Si to the respective electrodes, where electron travels through g-$C_3N_4$ while hole travels through p-Si, giving rise to a steady photocurrent in the external circuit without any external energy supply (Figure 8c).

In summary, we have developed a multifunctional hybrid photodetector based on the g-$C_3N_4$/p-Si heterojunction. The built-in field at the n-p junction allows the device to operate over ultrabroadband region from 250 to 1650 nm in self-powered mode. Under zero bias, the device shows high photosensitivity, responsivity, detectivity, and fast response speed at low-intensity light illuminations. More importantly, we demonstrate for the first time binary photoswitching behavior over a wide region from 250 to 1350 nm at small bias in a single device. The high-performance photodetector, with the advantages of easy fabrication process and nobel binary behavior makes it suitable for light-based binary communications, optoelectronic interconnects for fast power-efficient computing and weak signal detections. From these observations, it is revealed that ultrathin g-$C_3N_4$ nanosheets might potentially be also useful to improve the performance of Si solar cells significantly using the proposed technique.

**Experimental Section**

*Preparation of g-C$_3$N$_4$ nanosheets*: The as-prepared g-C$_3$N$_4$ yellow powder was commercially procured from Carbodeon Ltd Oy. The ultrathin g-C$_3$N$_4$ nanosheets were exfoliated from the bulk powder using ultrasonication method. 60 mg of bulk g-C$_3$N$_4$ was dispersed in 60 ml of ethanol (Alfa Aesar, 99% pure), and then the solution was sonicated in bath ultrasound sonicator for 2, 6 and 12 h. To maintain the temperature of solution during sonication, the water was changed at regular time intervals. Finally, the resulting sample solutions after exfoliation were labelled as gCN-2 (2 h), gCN-6 (6 h) and gCN-12 (12 h).

*Characterization of exfoliated nanosheets:* The surface morphology of as-prepared exfoliated nanosheets and the fabricated device was characterized by scanning electron microscopy and atomic force microscopy. A UV-Vis-NIR spectrometer (Agilent CARY5000) was used to analyse the light absorption of the samples. The XPS spectra of the samples were studied by using a microfocused monochromatic Al K$_\alpha$ X-ray source (1486.7 eV) and a seven channeltron hemispherical electron energy analyser (EA 125), under ultra-high vacuum conditions. The binding energy of the spectra were calibrated by considering the C1s peak position at 284.5 eV.

*Device Fabrication:* To prepare g-C$_3$N$_4$/Si heterojunction photodetector, a highly doped 5 mm × 5 mm, p-type silicon (100) wafer of resistivity 1-5 × 10$^{-4}$ Ωcm was used as the substrate. Prior to deposition of gCN-12, the p-Si substrates were degreased using isopropanol (IPA) and acetone. Subsequently, HF acid (49%) was drop-casted on p-Si for a fixed interval to etch the top silicon dioxide surface. Immediately, after this step, the sample was thoroughly rinsed with DI water to remove any residual acid. Thereafter, gCN-12 solution was drop-casted on the etched p-Si surface for fixed time interval. Afterwards, HF was again used to firstly remove the loosely adhered nanosheets and secondly for underetching into Si. This helped to ensure that the remaining nanosheets were ultrathin in dimensions and were properly adhered to the Si substrate providing a good electrical conductivity. It is well established that HF does not react with the g-$C_3N_4$.[43] Finally, the sample was thoroughly rinsed in DI water and blow-dried using dry nitrogen gas. This method yielded deposition of nanosheets throughout the substrates. Metal contacts of dimensions ~1 mm$^2$ were made by using silver paste on top of g-$C_3N_4$ and bottom p-Si surfaces to provide electrical connections. Similarly, control samples were prepared with top and bottom electrodes using silver paste on bare etched p-Si.

*Device Measurement:* Steady-state photoconductivity (PC) measurements were performed at room temperature. The PC set-up constitutes Xenon and Quartz-Halogen lamps with a TMc 300 monochromator and a source meter (Keithley 2635B). The light intensity of the lamps was measured by using a calibrated Silicon (Si-DH) photodetector. In order to measure the speed of the device, the monochromatic light was modulated by a mechanical chopper and signal was measured by a high-speed digital storage oscilloscope (DSO).

**Supporting Information**
Supporting information is available.

**Acknowledgements**
The authors are thankful to the Dr. D. K. Aswal, Director & Dr. V. N. Ojha, Head of the department, CSIR-NPL for constant encouragement. The authors would like to acknowledge Mr. Sandeep Kumar, CSIR-NPL for AFM and Mr. Shib Shankar Singha, Bose Institute for PL measurements. N. P. thanks to UGC, India for fellowship towards her Ph. D program. GK and SPK acknowledge the DST Start-Up Research Grant (Young Scientists), SERB Project file number YSS/2014/000803, India for financial support.